# A quantum mechanical analysis of the light-harvesting complex 2 from purple photosynthetic bacteria. Insights into the electrostatic effects of transmembrane helices

**Fabio Pichierri***

*G-COE Laboratory, Department of Applied Chemistry, Graduate School of Engineering, Tohoku University, Aoba-yama 6-6-07, Sendai 980-8579, Japan*

## Abstract

We perform a quantum mechanical study of the peptides that are part of the LH2 complex from *Rhodopseudomonas acidophila*, a non-sulfur purple bacteria that has the ability of producing chemical energy from photosynthesis. The electronic structure calculations indicate that the transmembrane helices of these peptides are characterized by dipole moments with a magnitude of ~150 D. When the full nonamer assembly made of eighteen peptides is considered, then a macrodipole of magnitude 704 D is built up from the vector sum of each monomer dipole. The macrodipole is oriented normal to the membrane plane and with the positive tip toward the cytoplasm thereby indicating that the electronic charge of the protein scaffold is polarized toward the periplasm. The results obtained here suggest that the asymmetric charge distribution of the protein scaffold contributes an anisotropic electrostatic environment which differentiates the absorption properties of the bacteriochlorophyll pigments, B800 and B850, embedded in the LH2 complex.

* Corresponding author. Tel. & Fax: +81-22-795-4132

*E-mail address:* fabio@che.tohoku.ac.jp (F. Pichierri)

## 1. Introduction

Photosynthetic bacteria (PB) have the ability to harvest sunlight by employing highly sophisticated protein-pigment complexes which are embedded in their membranes. Non-sulfur purple bacteria such as *Rhodopseudomonas sphaeroides* possess a pair of light-harvesting (LH) complexes, LH1 and LH2, which are coupled to a reaction centre (RC) as shown in the cartoon of Figure 1. The peripheral LH2 complex functions as an antenna by capturing photons with the aid of a lower ring made of nine bacteriochlorophyll molecules (B800) with absorption maxima around 800 nm. The radiation energy is then transferred to the upper ring made of 18 bacteriochlorophyll molecules (B850) which are sandwiched between the inner ring and outer ring of α and β peptides, respectively, which represent the backbone of LH2. Here the radiation can remain for about 10 ps. The energy is subsequently transferred to a ring of 36 bacteriochlorophyll molecules (B880) that belongs to LH1 and from here to the nearby bacteriochlorophyll molecules of the reaction centre (RC). Within RC the photons are transformed into chemical energy by oxidation of ubiquinone (UBQ) to ubiquinol (UBQH2). This reaction is coupled to electron transport processes that involve the metalloenzymes cytochrome b/$c_1$ and cytochrome c and generate a proton gradient across the membrane. The enzyme adenosine triphosphate (ATP) synthase, also embedded in the membrane, utilizes this proton gradient to synthesize ATP. This is how non-sulfur purple bacteria convert photon energy into chemical energy.

The atomic structure of the LH2 complexes from two purple bacteria, *Rhodopseudomonas acidophila* and *Rhodospirillum molischianum* have been characterized by x-ray crystallography. Although there are many similarities between their components, the LH2 complex of the former is a nonamer ($\alpha_9\beta_9$) whereas that of the latter is an octamer ($\alpha_8\beta_8$). Also, the crystal structure of the RC-LH1 complex from *Rhodopseudomonas palustris* was characterized at 4.8 Å resolution (Roszak et al., 2003). These important structural studies encouraged different research groups to perform detailed computational studies of the spectroscopic properties of the pigments that are embedded in the above complexes (Damjanović et al., 1999, Linnanto and Korppi-Tommola, 2004).

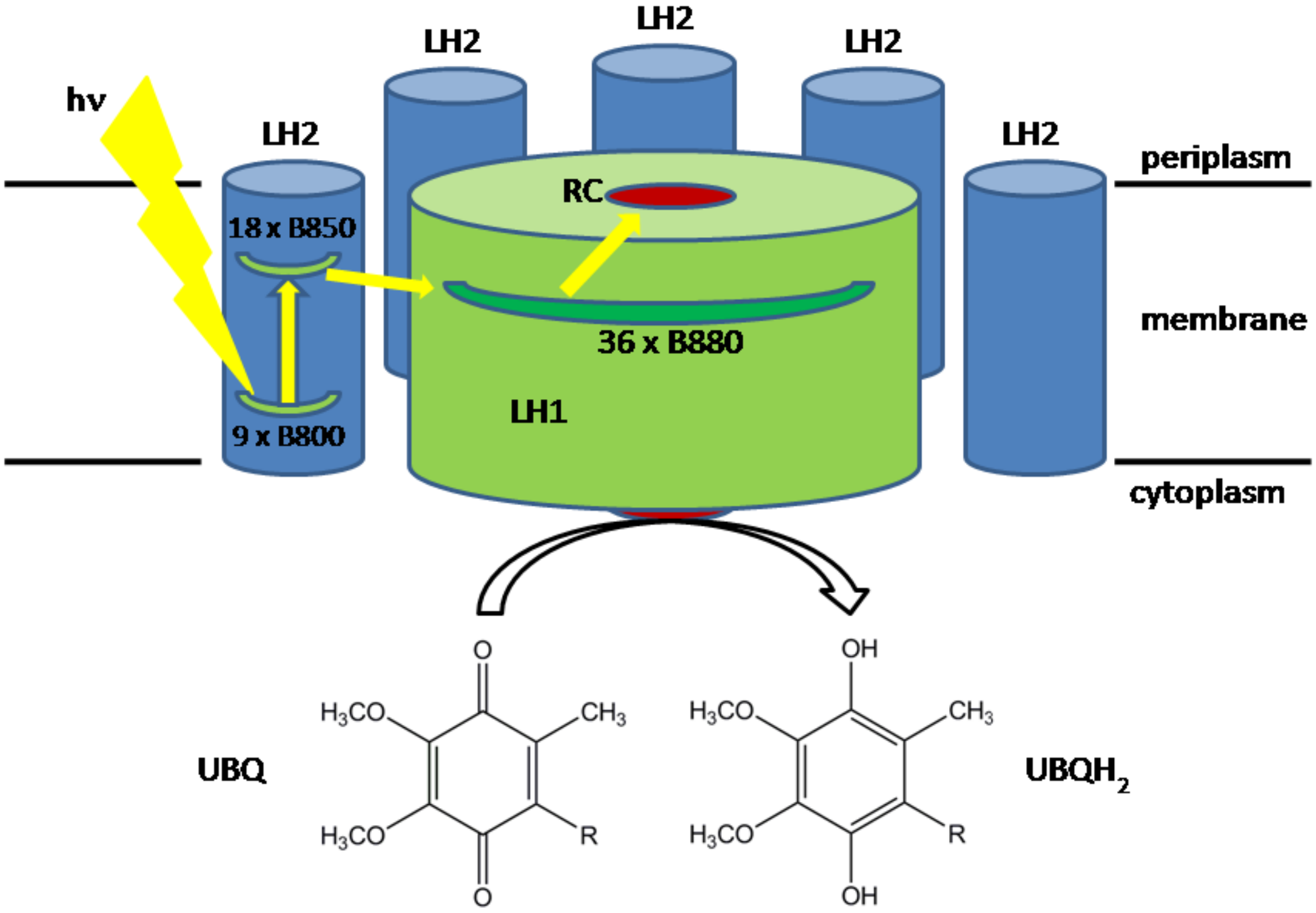

**Figure 1.** Schematic view of the photosynthetic system from non-sulfur purple bacteria. LH1 and LH2 are the light-harvesting complexes 1 and 2, respectively, while RC is the reaction centre. LH1 is provided with a ring of 36 BhCl while LH2 possesses two rings each made of 9 and 18 BhCl molecules, respectively. Reduction of the coenzyme ubiquinone (UBQ) to ubiquinol (UBQH) takes place in the RC (R represents an isoprenyl chain). Redrawn and modified from Nicholls and Ferguson (Nicholls and Ferguson, 2002).

A number of theoretical studies have shown that environmental effects from the molecules that surround the pigments in the complex supramolecular assembly of LH2 play an important role in determining the nature of the excited states associated to bacteriochlorophylls and carotenoids. In this regard, Pullerits and coworkers (He et al., 2002) have computed the transition energies of B800 while taking into account the local electrostatic environment of the protein described with an atomic charge field. It was observed that both the axial ligand to Mg(II) and an H-bonded Arg residue contributed to differences of ~25 nm in the transition energies of B800. We expect that the inclusion of additional amino acid residues may improve the results of excited states calculations. Increasing the size of the atomic model, however, increases dramatically the computational resources necessary to perform accurate electronic structure calculations. It is therefore important to evaluate the average electrostatic effects exerted by the protein scaffold on the bacteriochlorophylls so as to include them in future theoretical models of bacterial photosynthesis.

Here we perform a quantum mechanical analysis of the ground-state electronic structure of LH2 complex and its separated components. In particular, we want to quantify the magnitude of the helix dipoles associated to the α and β chains of LH2 so as to assess the magnitude of the electric field exerted by the entire protein scaffold on the bacteriochlorophyll rings. It is known that helix dipoles may have magnitude of up to one hundred debyes as they are built up from the vectorial sum of peptide bond dipoles (~1.85 D) along with the charged amino acid chains and C- and N-termini. Moreover, when two or more α-helices combine to form a biological assembly, then the vectorial sum of helix dipoles can produce macrodipoles with an associated electric field. In this regard, in a recent computational study (Pichierri, 2010) on the full length KcsA potassium channel, a homotetramer provided with four long α-helices, the author computed a total dipole moment of 403 D which indicates that the electronic charge of this channel is polarized toward the outside of the cell.

## 2. Computational methods and atomic models

Performing quantum mechanical calculations on large biomolecules is a rather challenging task which, however, has become feasible thanks to the development of novel algorithms and methodologies in the field of molecular quantum mechanics (quantum chemistry) in parallel to the increasing power of modern computers. Stewart developed a novel algorithm (called Mozyme) which employs localized molecular orbitals (LMOs) rather than conventional MOs by using which it is possible to perform calculations on large biomolecules that scale linearly with respect to the size of the system (Stewart, 1996). The most important feature of his algorithm is that of employing the Lewis structure of the molecule (i.e. lone-pairs, $\sigma$-type and $\pi$-type bonds) as the initial guess for the density matrix. From here onwards, the Fock matrix is computed until self-consistency is achieved.

Stewart's algorithm is implemented in the MOPAC2009 software package (Stewart, 2008) which we employ here for investigating the electronic structure of the $\alpha$ and $\beta$ peptides of LH2. This package implements several semiempirical NDDO methods among which the novel PM6 parameterization has been successfully tested on several proteins (Stewart 2007, Stewart 2009). Because the transmembrane helices of the $\alpha$ and $\beta$ peptides are immersed inside the membrane, the present calculations include the effect of the membrane environment which was modeled with the conductor-like screening model, COSMO (Klamt and Schüürman, 1993). A value of $\varepsilon$=9.0 has been used for the dielectric constant.

The atomic model utilized herein was obtained from the crystallographic coordinates (PDB accession code 1NKZ) of the LH2 complex from *Rhodopseudomonas acidophila* determined at 2.0 Å resolution and 100 K (Papiz et al., 2003). The atoms of the pigments (bacteriochlorophylls and carotenoids) as well as those of the solvent were removed and hydrogen atoms were added to saturate the open valences of heteroatoms (C, N, O, S). Quantum mechanical calculations were performed both on the $\alpha_1\beta_1$ monomer and on the $\alpha_9\beta_9$ nonamer, both shown in Figure 2a. In Figure 2 one can see that the bacteriochlorophylls B800 and B850 form an oligomer with the same nine-fold symmetry as that of the protein chains.

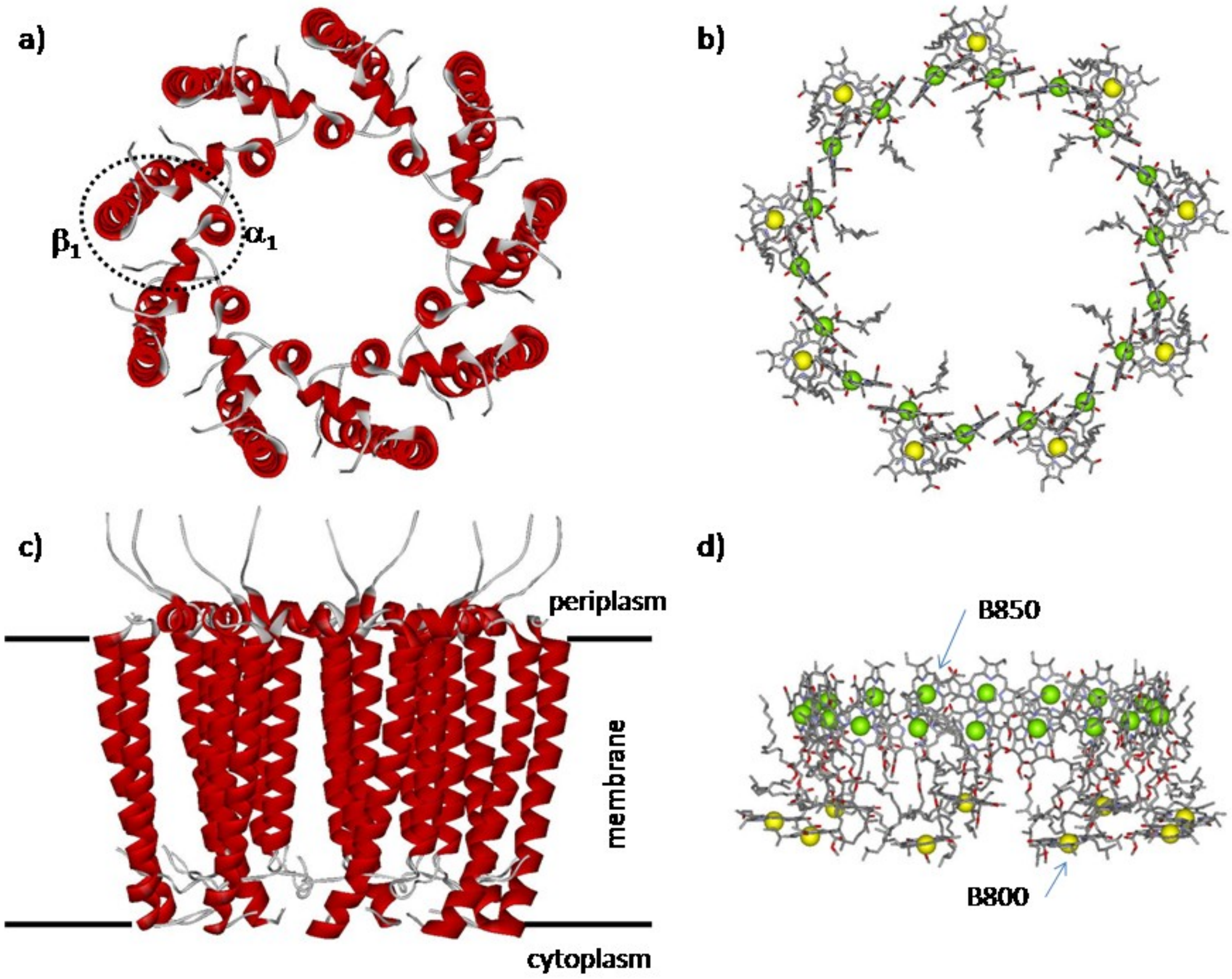

**Figure 2.** Components of the LH2 complex: (a) top view and (c) side view of $\alpha_9\beta_9$ with the transmembrane helices spanning the membrane; (b) top view and (d) side view of the bacteriochlorophyll units with their $Mg^{2+}$ atoms represented as yellow (B800) and green (B850) balls. The carotenoids are not shown in this figure.

## 3. Electronic structure of the α and β peptides of LH2

Our first task is that of determining the Lewis structure of the $\alpha_1$ and $\beta_1$ peptides of LH2. To this end, it is useful first analyzing the primary structure of these two chains with the aid of Figure 3. Chain $\alpha_1$ is composed of 53 amino acids of which four are positively charged (Lys5, His37, Lys50 and Lys51). The last residue, Ala53, is connected to the negatively-charged C-terminal which results in a net charge of +3 for this chain. Interestingly, the α peptide does not have a positively-charged N-terminal like most proteins but possesses an N-carboxymethionine residue which coordinates a Mg(II) ion of B800. In a previous structural study (Prince 1997) of the LH2 complex, an acetyl group was modeled at this site. The structure determined at 2.0 Å resolution, however, indicated the presence of a N-carboxymethionine residue but the protonation state of this group (COOH or $COO^{-}$) was not established (Papiz et al, 2003). In our atomic model of the $\alpha_1$ chain, we consider a neutral N-carboxymethionine group although a negatively-charged carboxylate group could exist when the peptide binds the metal centre.

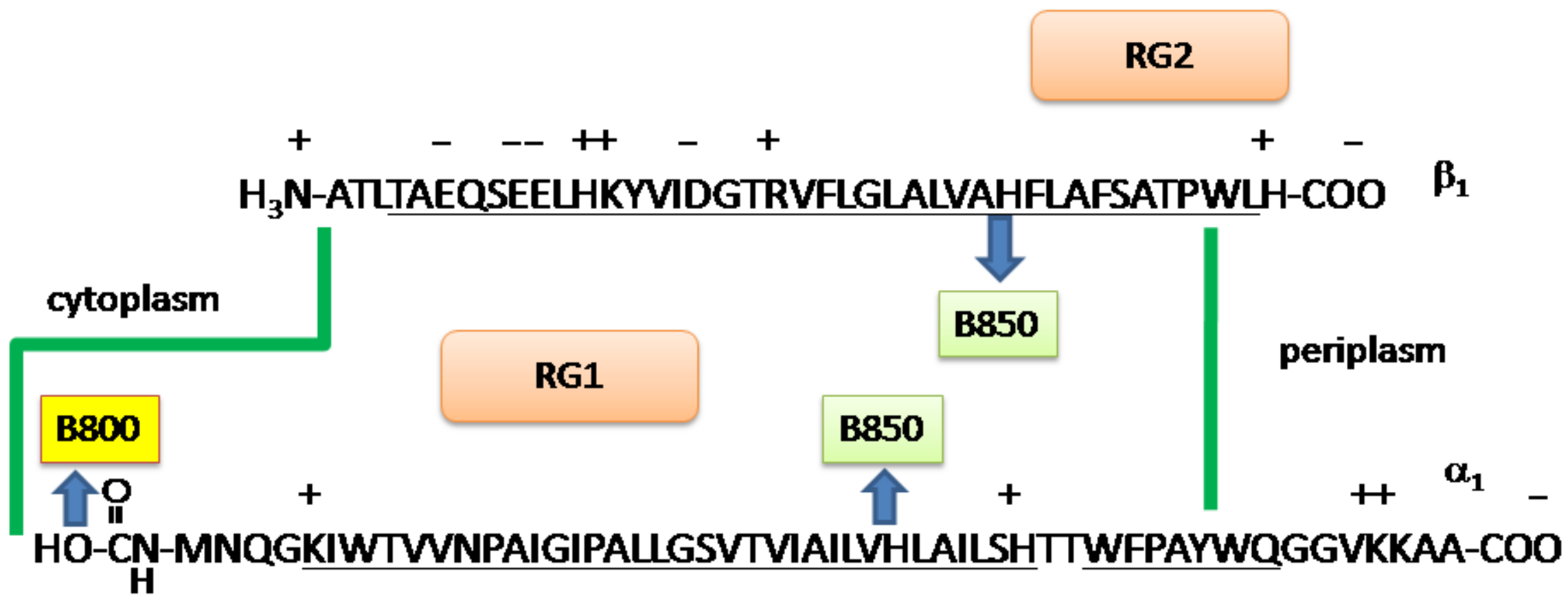

**Figure 3.** Primary structure of the $\alpha_1$ and $\beta_1$ peptides in the LH2 complex. The position of B800, B850 and the rhodopin glucosides (RG1 and RG2) within the αβ complex is also shown.

The His31 (neutral) residue coordinates the second B800 molecule. The secondary structure of the $\alpha$ chain is made of two $\alpha$-helices (underlined residues), a long one from Lys5 to His37 and a short one from Trp40 to Gln46, which are separated by a small kink located at Thr38 and Thr39.

The $\beta_1$ chain of LH2, on the other hand, is made of 41 residues, four of which are positively-charged (His12, Lys13, Arg20, His41) and three of which are negatively charged (Glu6, Glu9, Glu10, Asp17). These charged residues together with the positively-charged N-terminal and the negatively-charged C-terminal contribute a neutral $\beta$ chain. One nitrogen atom of the neutral His30 residue is coordinated to the Mg(II) centre of B850 whose molecular plane lies perpendicular to the membrane, as shown in Figure 2d.

The Lewis structure of the $\alpha_1$ peptide (54 amide bonds) is made of 840 $\sigma$-type bonds, 201 lone-pairs, and 79 $\pi$-type bonds. After achieving self-consistency, the resulting electronic structure is made of 1120 filled levels (occupied orbitals) and the computed HOMO-LUMO gap corresponds to 7.767 eV. The Lewis structure of the $\beta_1$ peptide (40 amide bonds) is made of 656 $\sigma$-type bonds, 172 lone-pairs, and 66 $\pi$-type bonds. The electronic structure obtained after achieving self-consistency is made of 894 filled levels and the corresponding HOMO-LUMO gap is 7.441 eV. This value is close to that computed for the $\alpha_1$ chain (hereafter we drop the indices 1 and 2).

The frontier orbitals of $\alpha$ and $\beta$ peptides are depicted in Figures 4 and 5, respectively. The three highest-occupied MOs of the $\beta$ peptide are localized on three different tryptophane residues: HOMO on Trp45, HOMO-1 on Trp7, and HOMO-2 on Trp40. In other proteins as well, we found that the tryptophane residue bears the filled levels that are higher in energy (Pichierri, 2004 and 2006). The HOMO, on the other hand, is found to be localized on His37. Hence, LUMO is spatially close to HOMO (~16 Å) and HOMO-1 (~11 Å) although these residues do not interact with each other.

As for the frontier orbitals of chain β, both HOMO and HOMO-1 are localized on Trp39 while HOMO-2 is found on Phe34. LUMO is quite far from these MOs being localized on His12. Taken together these results indicate that the highest-occupied MOs of both chains are localized on residues which are located near the periplasmic side of the membrane where B850 and RG2 are also located. Whether an interaction of the HOMOs with these two moieties has a functional role in the photosynthetic process of LH2 is an interesting issue which should be investigated in future computational studies.

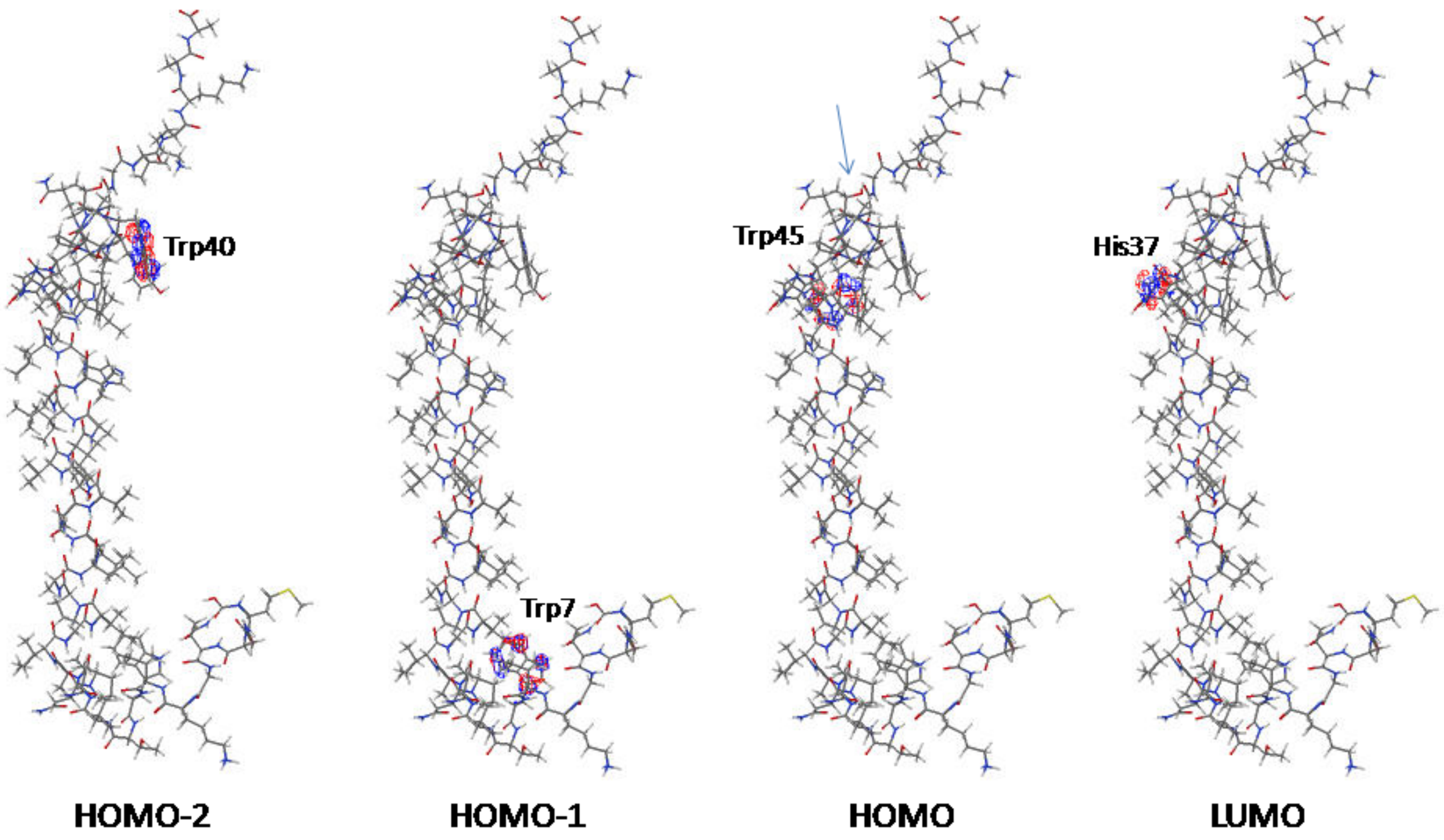

**Figure 4.** Frontier orbitals of the $\alpha_1$ chain.

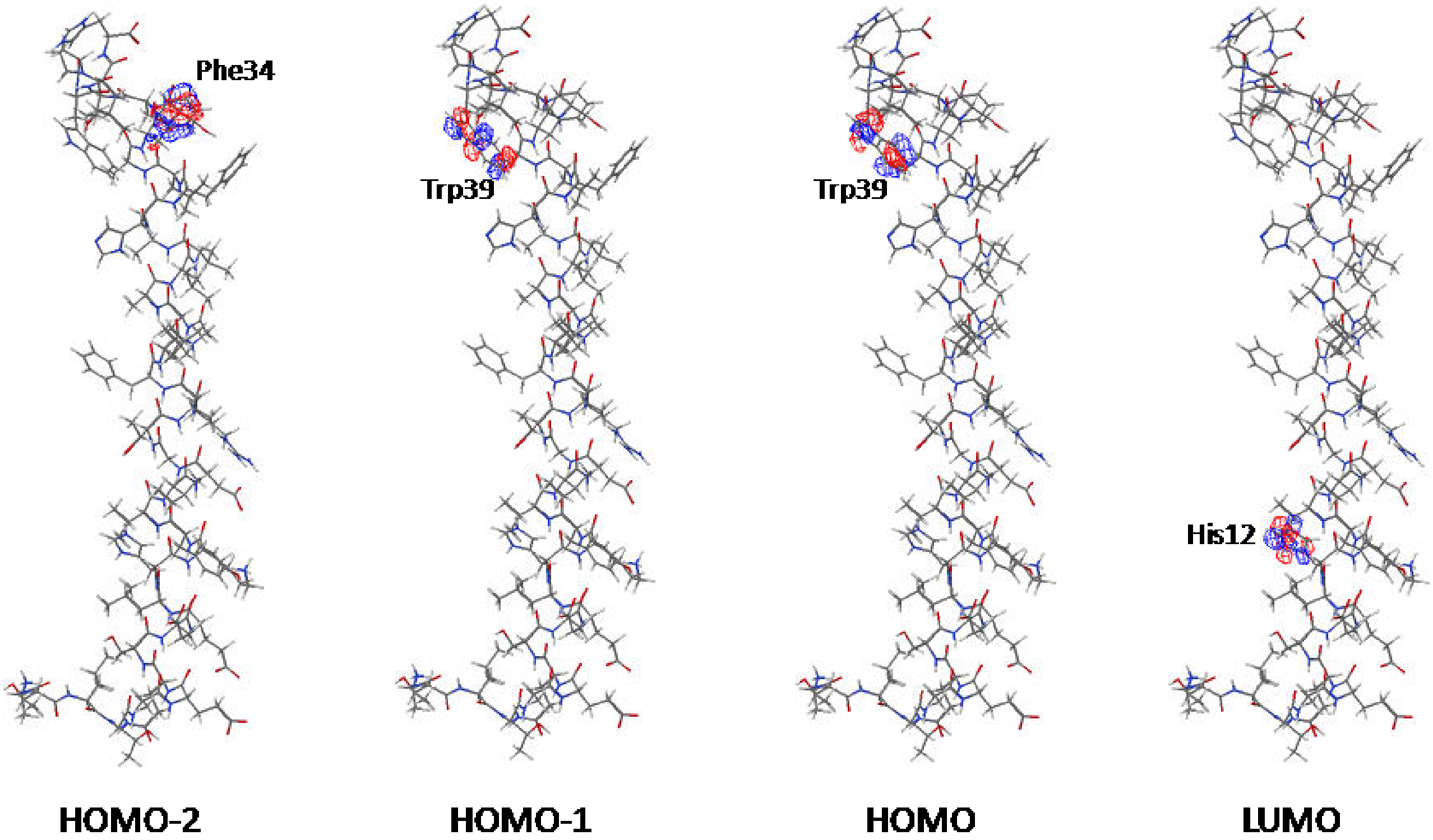

**Figure 5.** Frontier orbitals of the $\beta_1$ chain.

## 4. Dipole moments of LH2 components

Figure 6 shows the computed dipole moments of the $\alpha_1$ chain, $\beta_1$ chain, and $\alpha_1\beta_1$ complex of LH2. Interestingly, the dipole moments of $\alpha_1$ and $\beta_1$ chains are close in magnitude, being 147 D and 149 D, respectively. Also, we notice that the dipole moment vectors of these peptides form angles of 50° and 40° with the corresponding transmembrane helices. Once the $\alpha_1$ and $\beta_1$ chains are assembled to form the $\alpha_1\beta_1$ complex, the dipole moment vectors sum up to yield a dipole moment of magnitude 163 D. It is interesting to compare the relative position of the bacteriochlorophyll pigments with respect to the orientation of the dipole moment vector of $\alpha_1\beta_1$ complex. As can be seen from the drawing on the far right side of Figure 6, the Mg ions of B850 (green balls) are located in a region of space that is close to the negative end of the dipole vector while the Mg ion of B850 is located near the positive tip of the vector. This result indicates that the pigments within the LH2 complex are subjected to different electric fields which are likely to affect their characteristic absorption energy.

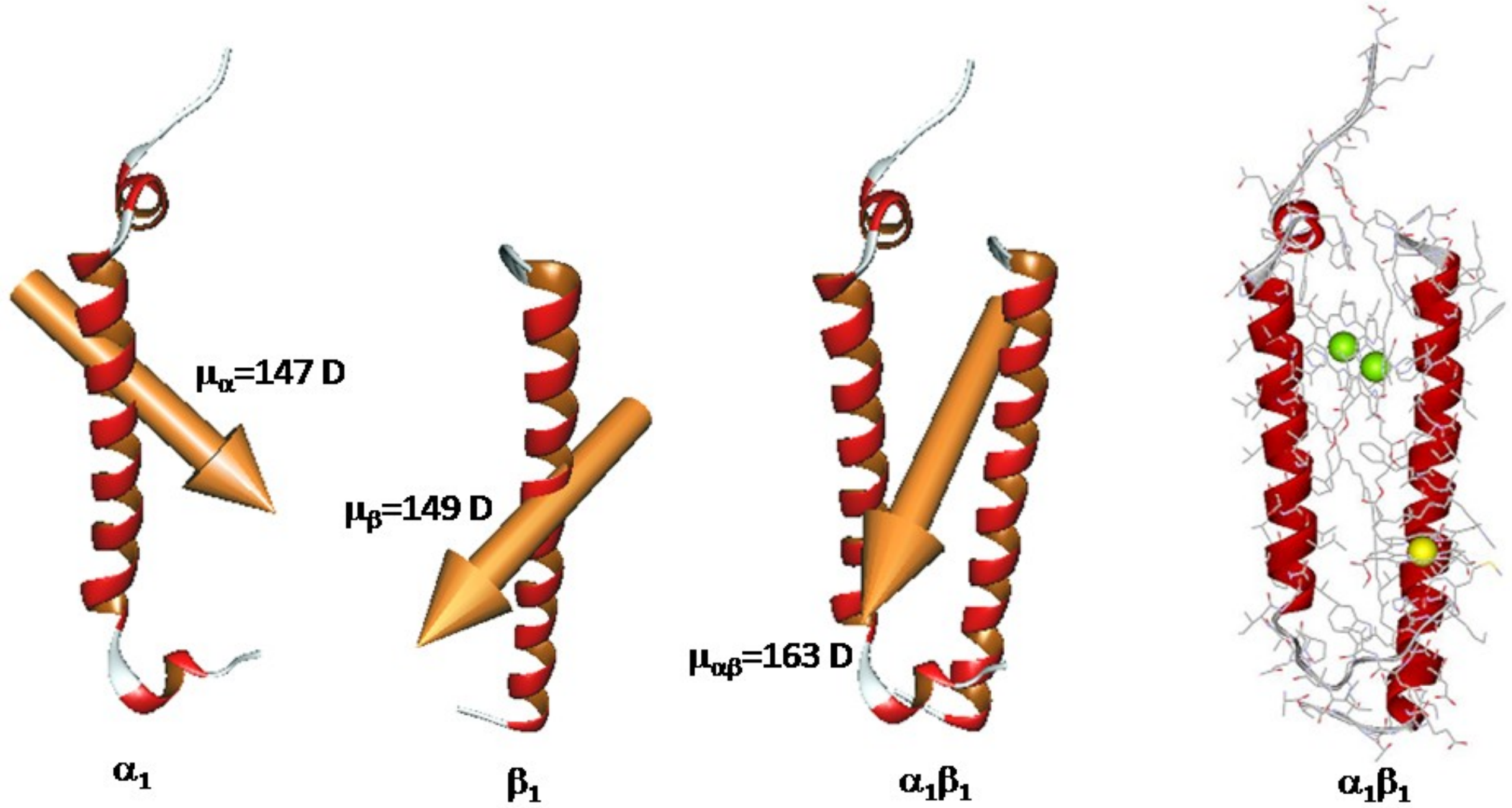

**Figure 6.** Computed dipole moments of $\alpha_1$ chain, $\beta_1$ chain, and $\alpha_1\beta_1$ chains. The structure on the far right shows the position of the $Mg^{2+}$ atoms of B800 (yellow ball) and B850 (green balls).

When nine $\alpha_1\beta_1$ chains are assembled to form the $\alpha_1\beta_1$ complex, a dipole moment with a magnitude of 704 D is built up, as shown in Figure 7. Further, the dipole moment vector is oriented normal to the membrane plane and with the positive tip directed towards the cytoplasm. This result indicates that the overall electronic charge density of the transmembrane helices is polarized toward the periplasm. Interestingly, we found that also the macrodipole of full-length KcsA potassium channel is oriented normal to the membrane plane while the electronic charge is polarized toward the outside of the cell membrane (Pichierri, 2010).

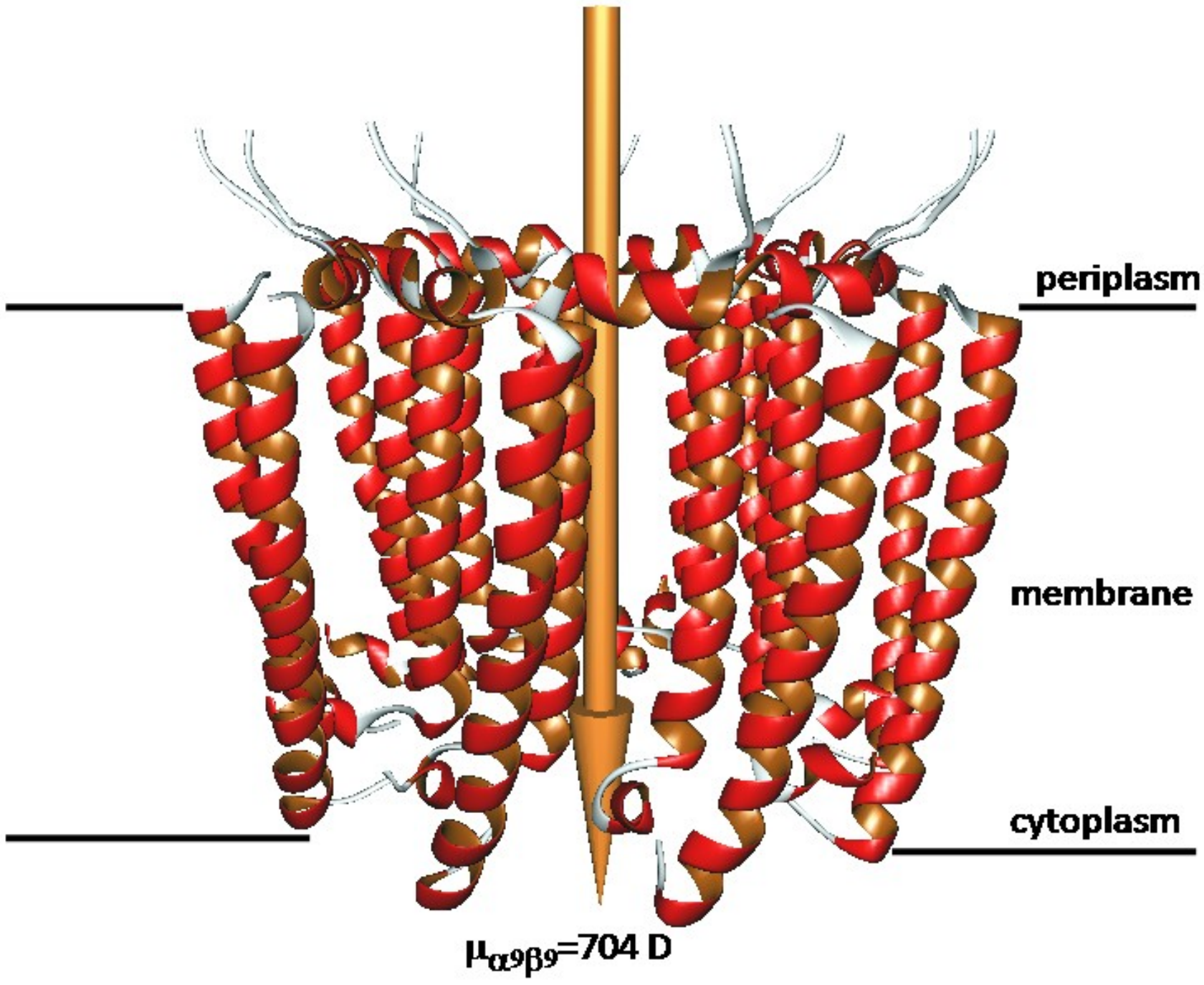

**Figure 7.** Macrodipole computed for the $\alpha_9\beta_9$ complex ($\alpha$ and $\beta$ chains only) of LH2.

## 5. Summary and conclusions

In summary, we performed a series of electronic structure calculations on the peptides that are part of the LH2 complex from *Rhodopseudomonas acidophila*. The results of the quantum mechanical calculations indicate that both $\alpha$ and $\beta$ peptides bear dipole moments with a magnitude of about 150 D. The vector sum of these dipoles generates a dipole moment of magnitude 163 D which characterizes the $\alpha_1\beta_1$ monomer and is responsible for creating an anisotropic electrostatic environment around bacteriochlorophylls B800 and B850. Furthermore, the $\alpha_9\beta_9$ complex possesses a macrodipole of magnitude 704 D which is oriented normal to the membrane plane and is responsible for polarizing the electronic charge towards the periplasm. We expect these results will contribute to better quantify the environmental effects of the protein scaffold that operate on the LH2 pigments.